# Monte Carlo Simulation of Electron Transport in Degenerate Semiconductors

Mona Zebarjadi[1], Ceyhun Bulutay[2], Keivan Esfarjani[3] and Ali Shakouri[1]

1-Department of Electrical Engineering, University of California, Santa Cruz, CA 95064
2-Department of Physics, Bilkent University, Ankara, 06800, Turkey
3-Department of Physics, University of California, Santa Cruz, CA 95064

A modified algorithm is proposed to include Pauli exclusion principle in Monte-Carlo simulations. This algorithm has significant advantages to implement in terms of simplicity, speed and memory storage. We show that even in moderately high applied fields, one can estimate electronic distribution with a shifted Fermi sphere without introducing significant errors. Furthermore, the free-flights must be coupled to state availability constraints; this leads to substantial decrease in carrier heating at high fields. We give the correct definition for electronic temperature and show that in high applied fields, the quasi Fermi level is valley dependent. The effect of including Pauli exclusion principle on the band profile; electronic temperature and quasi Fermi level for inhomogeneous case of a single barrier heterostructure is illustrated.

Many of today's interesting microelectronic devices are working in high doping concentrations up to $10^{20}$ cm$^{-3}$. Degenerate semiconductors are important for thermoelectric and thermionic energy conversion devices and they are also used in the highly doped source/drain regions of advanced transistors. At high carrier densities, the electron distribution is highly affected by its fermionic nature. Pauli exclusion principle (PEP) prohibits electrons from occupying the same state and thus pushes them to higher energy states or it can make certain scattering processes forbidden, or even block ballistic carrier motion. Therefore, it needs to be included in the theoretical analysis of electron transport in degenerate semiconductor-based structures.

If the electron wavelength is smaller than the characteristic lengths present in the structure, the Boltzmann transport equation (BTE) can be used as an appropriate governing equation for the device. The factor (1-*f*) in BTE, which indicates the probability of the final state to be unoccupied, is present in the scattering term as a result of PEP.

$$\frac{\partial f(\vec{r},\vec{k},t)}{\partial t} = -\vec{v}.\vec{\nabla}f(\vec{r},\vec{k},t) - \vec{F}/\hbar.\vec{\nabla}_k f(\vec{r},\vec{k},t) +$$

$$\sum_{k'}\{f(\vec{r},\vec{k'},t)W(\vec{k'},\vec{k})(1-f(\vec{r},\vec{k},t)) - f(\vec{r},\vec{k},t)W(\vec{k},\vec{k'})(1-f(\vec{r},\vec{k'},t))\}$$

Here *f* is the non-equilibrium distribution function, *F*, is the applied electric field, $\vec{v}$ is the group velocity and *W*, the scattering rate. The ensemble Monte Carlo (MC) simulation is accepted as a powerful numerical technique to solve BTE and is widely used to simulate transport in semiconductor devices. There have been some attempts to include PEP in MC simulations of degenerate semiconductors. All of these attempts have been based on the rejection method. At each scattering step, scattering to the final state is accepted with the probability of 1-*f* .However, in a MC simulation *f* is not a priori known.

 In a single-electron MC simulation Bosi and Jacoboni[1] suggested using $f(\vec{k},t)$ obtained from the simulation itself up to the time *t* at which the scattering is attempted. They evaluate $f(\vec{k},t)$ on a grid in *k*-space. Lugli and Ferry[2] (LF) proposed using the same method in ensemble MC simulation, but substituting the averaging over time by ensemble average at each time step. This algorithm is used by many other groups and is working well at high fields. But it is not suitable for low fields and highly degenerate cases. In low fields, it is reported that the algorithm leads to some unphysical results such as decrease of average electron energy with increase of electric field and values of electron distribution function exceeding one [3] . In order to improve the LF method, Borowik[3,4] suggested adding scattering-out terms into the simulation by introducing virtual scatterings in order to avoid *f* >1. With this method they were able to rebuild the Fermi Dirac distribution with a small deviation. We believe that in principle LF method is free from these artifacts at the cost of excessive *k*-space grid points, also implying a large number of simulated particles. Besides, if PEP is also checked after free- flights, the distribution function should not exceed one[5] . Another inconvenience of LF method is that this algorithm is suitable for a homogeneous steady state situation. In transient simulations and in inhomogeneous devices, tabulating the distribution function at various locations and times can require an unmanageable computation time and storage.

Fischetti *et al*[6] proposed to overcome this difficulty by approximating the local electron distribution by a quasi-equilibrium Fermi-Dirac distribution. It is simple and very cheap compared to the LF method and works well at low fields where one can use quasi equilibrium definitions to find the local temperature and chemical potential. Ref. 6 suggested approximating the electronic distribution by a quasi Fermi-Dirac distribution with the following definition.

$$f(E, \mu, T_{el}) = \frac{1}{\exp(\frac{E - \mu(\vec{r})}{k_B T_{el}(\vec{r})}) + 1} \quad , \qquad (1)$$

$$3/2 K_B T_{el}(\vec{r}) = \begin{cases} <E(\vec{r})> & non-degenerate \\ \mu(\vec{r}) & degenerate \end{cases}, \qquad (2)$$

Where $E$ is the energy of final state to which electrons tend to scatter, $\mu(r)$ is the local quasi Fermi level, $T_{el}(r)$ is the local electronic temperature and $<E(r)>$ is the local average kinetic energy of electrons. The above definition for electronic temperature is not correct at high concentrations when the Maxwell-Boltzmann distribution is not valid. The correct definition of electronic temperature, should converge to the lattice temperature at zero electric field, and should exceed it under an applied field. Moreover, under the applied bias, translational energy should be subtracted, because temperature is defined by fluctuations of electron velocities around its mean drift value. These considerations lead us to the following definitions:

$$f_\nu(E, \mu, T_{el}) = \frac{1}{\exp(\frac{E(|\vec{k} - \vec{k}_d^\nu(\vec{r})|) - \mu_\nu(\vec{r})}{k_B T_{el}^\nu(\vec{r})}) + 1} \qquad (3)$$

$$T_{el}^\nu(\vec{r}) = \frac{2}{3k_B}(<E(\vec{k} - \vec{k}_d^\nu(\vec{r}))> - <E_\nu>_0) + T_{lattice} \qquad (4)$$

Here $\nu$ is the valley index, $<E_\nu(\vec{r})>_0$ is the local average energy of electrons in equilibrium at zero electric field, which can be calculated analytically at each time step (see Eq. 6). $\vec{k}_d^\nu(\vec{r})$ is the local drift wave vector, which is the average wave vector of all the particles at position $r$ and in valley $\nu$. $\vec{k}$ is the   of the electron . For moderately high fields, separate Fermi spheres with different chemical potential and temperature need to be defined for each valley. Therefore, all quantities are valley dependent. Also in inhomogeneous cases such as heterostructures, we need to discretize the $x$ space and define the Fermi-level and the electronic temperature locally.

The implementation of the algorithm in the MC simulation is straightforward.

At each time step, drift wave vector ($\vec{k}_d$), average energy of electrons ($<E(\vec{k} - \vec{k}_d)>$) and local charge ($n_c$) are calculated.  Then electronic temperature, Fermi level and $<E>_0$ can be calculated using equations 4, 5 and 6 respectively.

$$n_c = \int\limits_o^\infty f(\varepsilon, \mu, T_{el})\, g(\varepsilon)\, d\varepsilon \quad , \tag{5}$$

$$< E(\vec{r}) >_0 = \int (\varepsilon - \varepsilon_c)\, f(\varepsilon, \mu(\vec{r}), T_{lattice})\, g(\varepsilon)\, d\varepsilon \quad , \tag{6}$$

Where $g(\varepsilon)$ is the density of states and $\varepsilon_c$ is the bottom of conduction band. These updated quantities will be used in Eq. 3 for scattering probabilities at the next iteration.

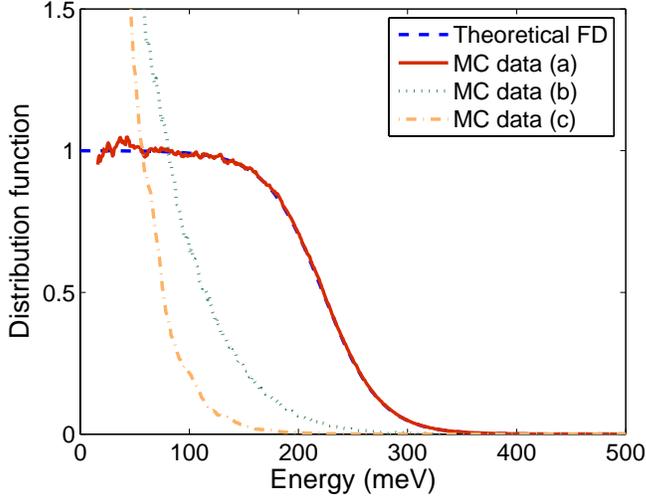

**FIG.1. Electron energy distribution function at equilibrium for GaAs at room temperature and the doping of $10^{19}$ cm$^{-3}$. Theoretical Fermi Dirac function is plotted for comparison. The rest are Monte Carlo simulation results implementing: (a) Present algorithm, (b) Algorithm of Ref.6 (Eqs. 1 and 2) and (c) Without Pauli exclusion principle.**

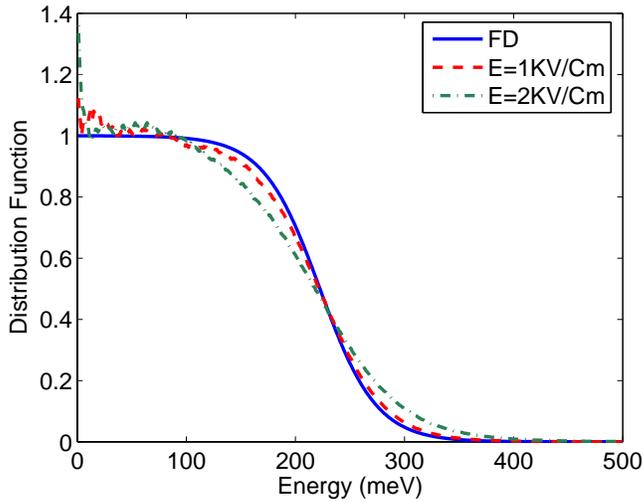

**FIG. 2. Electron energy distribution function for two different applied fields obtained from MC simulation. The simulation is done under the same condition as figure 1. Theoretical Fermi-Dirac function is plotted for comparison.**

Figure 1 shows the distribution at zero electric field obtained from the present algorithm in comparison with analytical Fermi-Dirac (FD) distribution. The deviation is negligible in most of the energy range. However, using definitions of Ref. 6 (Eqs. 1,2), the obtained final distribution is totally different from the FD distribution. Figure 2 shows the results of implementing the algorithm for two different low applied electric fields. As expected some of the electrons below Fermi level are pushed to higher energy levels producing thereby heating.

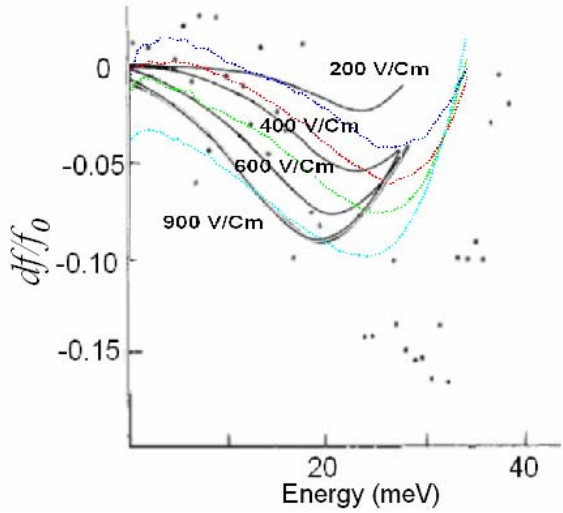

**FIG. 3. Relative change of the electronic distribution to the equilibrium(FD distribution) at 77 K temperature and the doping density of $5\times10^{17}$ cm$^{-3}$. Experimental results (solid curve) are after Ref. 8. MC data obtained from our simulation for the same applied electric fields are shown by dot lines. Previous MC simulation results of Ref 1. are shown by dots.**

Figure 3 shows a comparison between our algorithm and experimental results[7]. The experiment has been done at 77K on Te-doped GaAs. In low temperatures, impurity scattering is the dominant scattering mechanism. In this simulation, we have included both ionized and neutral impurity scattering. Neutral impurities are considered as hard spheres with the potential of 35 eV and radius of 2 Å. A binding Energy of 0.03 eV is considered for tellurium in GaAs[8]. Polar optical and acoustic phonons are also included, both as inelastic processes. Although there is a deviation compared to the experimental results, our results are closer to the experiment compared to the previous work of Ref. 1.

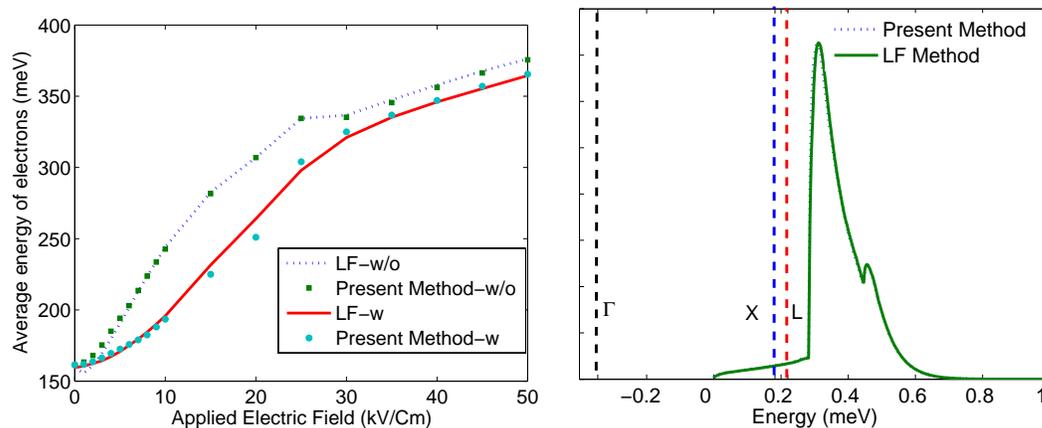

**FIG. 4. Comparison between LF method and the present method: a) Average energy of electrons versus applied electric field, with (w) and without (w/o) checking PEP after free-flights. b) Distribution function at 45 kV/cm obtained form the present method and the comparison with LF method. Fermi levels for Γ, L and X valleys are shown with dashed lines.**

In figure 4 we show a comparison between our suggested algorithm and the LF method. At very low fields, LF method gives nonphysical results as explained before due to insufficient number of electrons (100,000), but it is accurate at high fields. The agreement between the present method and the LF method suggests that the present algorithm works well even at moderately high fields of the order of several 10kV/cm. As a matter of fact, the agreement between the two methods subsists until at least 50 kV/cm, but our modeling (non-parabolic multi-valley band structure) becomes questionable at such fields for GaAs.   The figure is plotted for high doing of $10^{19}$ cm$^{-3}$. In such a high doping LF method gives reasonable results with at least 700,000 electrons (this large number of electrons is essential especially at low fields), however the present method is working well even with a sample of 10,000 electrons implying a reduction in CPU time and memory by a factor of 70! Moreover, even with the same number of electrons LF method uses 4.7 times the memory and it is 15% slower in comparison with the present method. These simulations were done on a PC with Pentium(R) 4 processor and 2.0 GB of memory.

It is noticeable that the agreement with LF method, at high fields is obtainable only if quasi Fermi level for each valley is defined separately. This is because at high fields we are far from equilibrium and electrons in different valleys do not equilibrate with each other. Another issue is that one needs to check PEP whenever there is a change in the state of the electrons; this includes both scatterings and free-flights. Although it is not a common practice to check PEP at free-flights[1-3, 5, 6], we noticed that this check affects the results (cf. Figure 4).

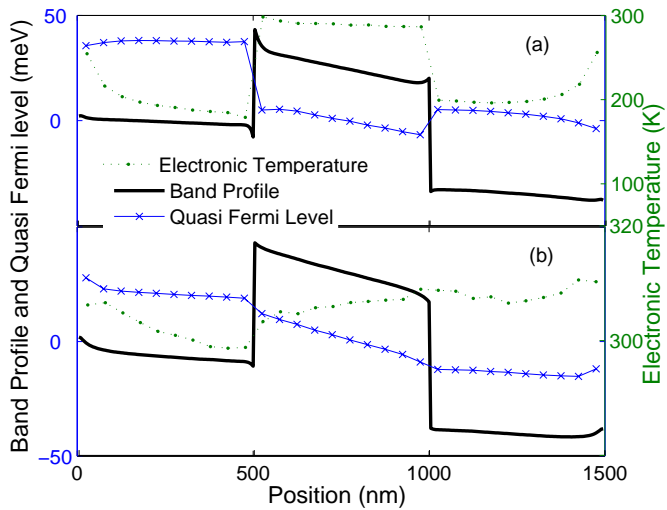

**FIG.5. Band Profile, quasi Fermi level and electronic temperature for InGaAs/InGaAsP/InGaAs heterostructure at room temperature under a low applied bias without (a) and with (b) PEP.**

Finally, we applied this approach to the inhomogeneous case of a single barrier heterostructure. In the absence of bias (equilibrium case) the quasi Fermi-level is constant in the whole device. Without considering PEP, however, distribution would not lead to a constant Fermi level. Figure 5 shows the electronic temperature and quasi Fermi level under a low applied bias, with and without applying PEP. The simulation is at room temperature. Without PEP, electrons are much colder than the lattice temperature especially in the contact layers where Fermi level is within the conduction band. In this case electrons would go from the barrier to the lower occupied states in the contacts and overpopulate them. This would lead to an artificial band bending which is shown in figure 5. By including degeneracy in the calculation, distribution leads to continuous Fermi level and electronic temperature is close to the lattice temperature. Cooling of electrons before the barrier and heating of them after the barrier can be explained as Peltier cooling and heating, and heating inside the barrier is a result of Joule heating. The transition between nonlinear thermionic emission cooling and linear transport is discussed in a recent publication[9]

In summary, we proposed an improved algorithm which handles degeneracy in highly doped semiconductors. We showed that the algorithm works well in homogenous and inhomogeneous cases and it requires much less time and memory storage compared to the other methods. This allows the treatment of inhomogeneous systems, which is almost impossible task with the LF method. Comparisons with analytical results at zero bias, and with other algorithms and also experimental data under applied bias were also presented. The effect of including PEP in a heterostructure on the band profile, electronic temperature and the quasi Fermi level was discussed.

M.Z. is thankful to Prof. Ravaioli for his helpful discussions. We also wish to acknowledge the support by ONR Thermionic Energy Conversion Center MURI.

---